\documentclass[aps,prd,superscriptaddress,twocolumn,nofootinbib]{revtex4-2}
\usepackage[utf8]{inputenc}
\usepackage{dsfont}
\usepackage{amsfonts}
\usepackage{amsmath}
\allowdisplaybreaks[4]        
\usepackage{amssymb}
\usepackage{euscript}     
\usepackage{braket}
\usepackage{starfont}
\usepackage{color,soul}         
\usepackage{tensor}        
\usepackage{amsthm}
\usepackage{graphicx}
\usepackage{slashed}
\usepackage{leftidx}
\usepackage{subfigure}
\usepackage{bbm}
\definecolor{outerspace}{rgb}{0.25, 0.29, 0.3}
\definecolor{scarlet}{rgb}{1.0, 0.13, 0.0}
\usepackage[header,title,page,titletoc]{appendix}  
\definecolor{princetonorange}{rgb}{1.0, 0.56, 0.0}
\definecolor{WildStrawberry}{rgb}{1.0, 0.26, 0.64}
\definecolor{rossocorsa}{rgb}{0.83, 0.0, 0.0}
\definecolor{navyblue}{rgb}{0.0, 0.0, 0.5}
\usepackage{float}
\usepackage[paper=letterpaper,margin=1in]{geometry}
\usepackage{mathtools}

\usepackage{amstext} 
\usepackage{array}   
\newcolumntype{L}{>{$}l<{$}} 
\usepackage{diagbox}
\usepackage{booktabs}
\usepackage{graphicx}
\usepackage{tabularx}
\usepackage{ragged2e}
\usepackage{siunitx}

\DeclareMathAlphabet{\pazocal}{OMS}{zplm}{m}{n}
\newcommand{\req}[1]{(\ref{#1})} 

\newcommand{\bea}{\begin{eqnarray}}

\newcommand{\eea}{\end{eqnarray}}
\newcommand{\ba}{\begin{eqnarray}}
\newcommand{\ea}{\end{eqnarray}}

\newcommand{\be}{\begin{equation}}
\newcommand{\ee}{\end{equation} }
\newcommand{\beqa}{\begin{eqnarray}}
\newcommand{\eeqa}{\end{eqnarray}}
\newcommand{\beqar}{\begin{eqnarray*}}
\newcommand{\eeqar}{\end{eqnarray*}}

\renewcommand{\req}[1]{(\ref{#1})}

\makeatletter
\newcommand{\dal}{\mathop{\mathpalette\dal@\relax}}
\newcommand{\dal@}[2]{%
  \begingroup
  \sbox\z@{$\m@th#1\square$}%
  \dimen0=\fontdimen8
    \ifx#1\displaystyle\textfont\else
    \ifx#1\textstyle\textfont\else
    \ifx#1\scriptstyle\scriptfont\else
    \scriptscriptfont\fi\fi\fi3
  \makebox[\wd\z@]{%
    \hbox to \ht\z@{%
      \vrule width \dimen0
      \kern-\dimen0
      \vbox to \ht\z@{
        \hrule height \dimen0 width \ht\z@
        \vss
        \hrule height 2\dimen0
      }%
      \kern-2.5\dimen0
      \vrule width 2.5\dimen0
    }%
  }%
  \endgroup
}
\makeatother

\usepackage{hyperref}
\hypersetup{
    colorlinks,
    citecolor=navyblue,
    filecolor=navyblue,
    linkcolor=navyblue,
    urlcolor=navyblue
}

\begin{document}

\title{Amplification of new physics in the quasinormal mode spectrum of highly-rotating black holes}

\author{Pablo A. Cano}
\email{pablocano@um.es}
\affiliation{Departamento de F\'isica, Universidad de Murcia, Campus de Espinardo, 30100 Murcia, Spain }
 \affiliation{Departament de F\'isica Qu\`antica i Astrof\'isica, Institut de Ci\`encies del Cosmos\\
 Universitat de Barcelona, Mart\'i i Franqu\`es 1, E-08028 Barcelona, Spain }
 
 \author{Marina David}
 \email{marina.david@kuleuven.be}
 \affiliation{Institute for Theoretical Physics, KU Leuven,
	Celestijnenlaan 200D, B-3001 Leuven, Belgium \vspace{0.1cm}}
 \affiliation{Leuven Gravity Institute, KU Leuven, Celestijnenlaan 200D, B-3001 Leuven, Belgium}
	
\author{Guido van der Velde}
\email{guidovandervelde@icc.ub.edu}
\affiliation{Departament de F\'isica Qu\`antica i Astrof\'isica, Institut de Ci\`encies del Cosmos\\
 Universitat de Barcelona, Mart\'i i Franqu\`es 1, E-08028 Barcelona, Spain }

\date{October 20, 2025}

\begin{abstract}
We show that perturbatively-small higher-derivative corrections to the Einstein-Hilbert action can lead to order-one modifications of the quasinormal mode spectrum of near-extremal Kerr black holes. The spectrum of such black holes contains zero-damping modes (ZDMs) and damped modes (DMs), with the latter only existing when the ratio $\mu=m/(l+1/2)$ is below a critical value $\bar{\mu}_{\rm cr}\approx 0.744$. Thus, this value represents a ``phase boundary'' that separates a region with both ZDMs and DMs and a region with only ZDMs. We find that the modes lying close to the phase boundary are very sensitive to modifications of GR, as their lifetimes receive corrections inversely proportional to their distance to the boundary. We link this growth of the corrections to a modification of the critical point $\bar{\mu}_{\rm cr}$, which can lead to a change in the number of DMs and produce order-one effects in the spectrum. We show that these large effects can take place in a regime in which the higher-derivative expansion remains under control. We also perform an exact analysis of the modification of the phase boundary for lower $(l,m)$ modes and pinpoint those that are most sensitive to corrections. Our results indicate that spectroscopy of highly-rotating black holes is by far the most powerful way to search for new physics in ringdown signals. 
\end{abstract}

\maketitle

Research on black holes has long fueled progress on our exploration of the gravitational interaction.
For decades, black holes have served as a theoretical laboratory to test ideas about a hypothetical theory of quantum gravity, or in general, about any new physics that may lie beyond the realm of classical general relativity (GR).  
With the discovery of gravitational waves \cite{LIGOScientific:2016aoc}, probing the nature of black holes experimentally has now become a real possibility. This opens up the opportunity to place constraints on some of our ideas on beyond-GR physics \cite{Berti:2018cxi,Berti:2018vdi,Barack:2018yly,Barausse:2020rsu}. While at first  this may look like a long shot, the first step towards this goal is to understand what the clear signatures --- the ``smoking guns'' --- of new physics are. This analysis may reveal where to look if we want to maximize our possibilities of detecting physics beyond GR.  And in this regard, near-extremal black holes could provide us with the best opportunity. 

Extremal black holes are characterized by possessing a degenerate event horizon, an infinitely long throat and zero surface gravity.  This requires that the black hole is charged and/or rotating, with the charge and angular momentum saturating a certain bound.  A near-extremal black hole is one that is close to saturating such bound and therefore has a small but non-vanishing Hawking temperature and a long but finite throat in the near-horizon region.  While extremal black holes cannot form by usual matter accretion\footnote{Extremal black hole formation is possible for certain types of matter though \cite{Kehle:2022uvc}.} \cite{Bardeen:1973gs,Israel:1986gqz}, near-extremal black holes, in particular, neutrally-charged highly-rotating black holes, could actually exist in our universe, so our interest in them is more than theoretical.  
These black holes possess intriguing features already at the level of classical GR. On the one hand, extremal black holes are prone to the Aretakis instability \cite{Aretakis:2012ei,Lucietti:2012sf}, by which the derivatives of fields at the event horizon blow up polynomially with time. On the other hand, although near-extremal ones are probably stable \cite{Hafner:2025nrv}, they possess long-lived modes that propagate on the throat of the black hole and that give rise to a very slow decay of perturbations \cite{Hod:2008zz,Hod:2009td,Hod:2012bw,Yang:2012pj,Yang:2013uba}. 
These properties make these black holes potentially sensitive to new physics beyond classical GR, and recent developments are providing evidence in this direction. 

For instance, it has been recently shown that quantum gravitational effects become important when black holes are close enough to extremality \cite{Iliesiu:2020qvm}.  This phenomenon is controlled by the Planck scale, with quantum fluctuations becoming relevant when the temperature is below the value $r_{+} T\lesssim \ell_{P}^2/r_{+}^2$, where $r_{+}$ is the horizon radius and $\ell_{P}$ is the Planck length. For black holes of astrophysical size, this requires being unrealistically close to extremality, so that we cannot hope to observe these effects. 

A different set of effects of new physics corresponds to classical corrections to GR, usually in the form of higher-derivative terms in the Einstein-Hilbert action. Unlike quantum effects, these classical corrections depend on an undetermined length scale of new physics, potentially much larger than the Planck length. Hence, they could lead to important effects on near-extremal black holes much before quantum effects become relevant.  
In this front, it was observed by  \cite{Horowitz:2023xyl} that higher-derivative corrections to GR can produce a divergence of tidal forces on the horizon of extremal black holes when these are perturbed by an external field.\footnote{A  similar phenomenon was earlier observed for AdS black holes in \cite{Horowitz:2022mly}.} Subsequent literature has further expanded this analysis \cite{Horowitz:2024dch,Cano:2024bhh} and has tried to clarify the meaning of those divergences \cite{Chen:2024sgx,Chen:2025sim, DelPorro:2025fiu}. Furthermore, some theories beyond GR can lead to even more extreme phenomena, with stationary black hole solutions becoming singular at extremality \cite{Kleihaus:2015aje, Chen:2018jed}. All these divergences (both in the background solution and in the perturbations) go away in the near-extremal case, but they imply that the corrections to GR grow unboundedly as extremality is approached. This would mean that the observation of a near-extremal black hole could allow us to put much stronger constraints on the scale of new physics. However, despite these inspiring results, there is still no clear understanding of the possible consequences of these findings for observables.
 
 In this paper, we identify a new phenomenon of an ``amplification of new physics'' that would apply to any small modification of GR and that concerns some of the most relevant observables of the theory: the quasinormal modes (QNMs) of black holes.  These are the characteristic vibrational modes of black holes and they govern the gravitational wave emission during the ringdown phase in black hole mergers \cite{Kokkotas:1999bd,Dreyer:2003bv,Berti:2009kk,Konoplya:2011qq,Berti:2025hly}.  Thus, they are key observables to test GR and modifications thereof via black hole spectroscopy \cite{Silva:2022srr,2023arXiv230501696F,Maselli:2023khq,Liu:2024atc,Maenaut:2024oci}.
In particular, we focus on the QNM spectrum of near-extremal Kerr black holes.

QNMs exist for a discrete set of complex frequencies whose imaginary part represents the inverse of the damping timescale. The QNM spectrum of a black hole is thus labeled by the harmonic numbers $l 
, m$ and by an overtone index $n=0,1,2, \ldots$ that orders the modes according to their damping time (with $n=0$ corresponding to the longest lived mode).
For Kerr black holes, the frequencies $\omega_{l m n}$ are also functions of the rotation parameter $a=J/M$, and the approach to extremality $a\to M$ turns out be very intriguing. As discovered by \cite{Yang:2012pj,Yang:2013uba}, the spectrum of near-extremal Kerr black holes  bifurcates into two families of modes: zero-damping modes (ZDMs), whose imaginary part vanishes in the extremal limit, and damped modes (DMs), for which it tends to a non-zero value. Thus, the former are infinitely long-lived in the extremal limit. 
As found by \cite{Yang:2012pj,Yang:2013uba}, ZDMs exist for all co-rotating modes with $m\ge 0$, while DMs only exist for a restricted set of values of the ratio 
\begin{equation}\label{mudef}
\mu=\frac{m}{L}\, ,
\end{equation}
where $L=l+1/2$. A computation based on the eikonal approximation $l\gg 1$ shows that DMs exist for $\mu$ below a critical value $\mu<\bar{\mu}_{\rm cr}\approx 0.744$. 
In this paper, we show that the modes lying very close to the phase boundary between DMs and ZDMs are very sensitive to modifications of GR, and that their lifetimes can receive arbitrarily large corrections. 

The main challenge in analyzing this question is describing the perturbations of black holes with arbitrary rotation in theories beyond GR.  Despite remarkable recent progress in this direction \cite{Cano:2023jbk,Chung:2024ira,Chung:2024vaf,Blazquez-Salcedo:2024oek,Cano:2024ezp,Khoo:2024agm,Blazquez-Salcedo:2024dur,Chung:2025gyg}, the case of very high rotation is in general out of reach.  An exception to this is the case of the quartic-curvature theory identified by \cite{Cano:2024wzo} as the only ``eikonal-isospectral'' effective field theory (EFT) extension of GR to eight derivatives. By using the properties of this theory, it has been recently shown by \cite{Cano:2025mht} how to obtain the QNMs of black holes of arbitrary rotation in the eikonal limit.  
Thus, we use this theory as an illustrative and relevant case to understand the corrections to QNMs near extremality. Our analysis will nevertheless reveal that the phenomenon we uncover is general and would apply to essentially any modification of GR.

The rest of the paper is organized as follows. 
We start by reviewing the QNM spectrum of near-extremal Kerr black holes in section~\ref{sec:QNMKerr}. Focusing on the eikonal modes, we show how ZDMs and DMs can be identified via the WKB method and obtain the condition for the phase boundary that separates both types of modes. We also give explicit expressions for the QNM frequencies near extremality and near the phase boundary. 
In section~\ref{sec:QNMHD}, we compute higher-derivative corrections to the QNMs. First we show that the perturbative correction to the imaginary part of the QNM frequencies blows up in the extremal limit for modes close to the phase boundary. Then, we also perform a non-perturbative analysis of the QNM frequencies, showing that the blow-up of the relative corrections is actually related to a modification of the phase boundary condition. As a consequence, the nature of some modes can change on account of the corrections, implying that the number of DMs can be increased or reduced.  
In section~\ref{sec:regimeofvalidity} we discuss whether this large change in the spectrum can take place within the regime of validity of the EFT, showing that it can indeed occur in a regime in which the higher-derivative expansion remains under control. 
While the previous sections apply to the eikonal limit, in section~\ref{sec:sensitivemodes} we perform an exact analysis of the modification of the phase boundary for low $l$ modes. We identify the values of $(l,m)$ most sensitive to corrections and provide an estimate for the size of corrections. 
We conclude in section~\ref{sec:discussion}.

\section{Quasinormal modes of near-extremal Kerr black holes}\label{sec:QNMKerr}
The Kerr metric is given by the line element
\begin{equation}
\begin{aligned}\label{eq:Kerr}
	ds^2 =& -\frac{\Delta}{\Sigma}\left(dt-a (1-x^2)d\phi\right)^2 + \frac{\Sigma}{\Delta}dr^2\\
	& + \frac{\Sigma}{1-x^2}dx^2 + \frac{1-x^2}{\Sigma}\left(adt-(r^2+a^2)d\phi\right)^2\, ,
\end{aligned}
\end{equation}
where $x=\cos\theta$ and
\begin{align} \label{eq:DeltaWsol}
	\Delta = r^2 - 2 M r + a^2, \quad \Sigma = r^2 + a^2 x^2\, .
\end{align}
Here, $M$ is the mass of the black hole and $a=J/M$ where $J$ is the angular momentum. For $|a|\le M$, \req{eq:Kerr} represents a black hole  with an event horizon located at $r_{+}=M+\sqrt{M^2-a^2}$. We will focus on the near-extremal regime, corresponding to
\begin{equation}
\epsilon\equiv 1-\frac{a}{M}\ll 1\, .
\end{equation}
In order to understand the main features of the spectrum of near-extremal Kerr black holes, it is enough to consider the eikonal approximation $l\gg 1$. In this regime, the behavior of perturbations is independent of the spin of the field, and can be universally described by the wave equation for a test scalar field $\nabla^2\Phi=0$. Separation of variables $\Phi= e^{-i\omega t+i m\phi} S_{l m}(x) \psi_{l m}(r)$ then yields the Teukolsky radial and angular equations. 

The angular equation reads
\begin{align}\notag
&\frac{d}{dx}\left[(1-x^2) \frac{dS_{l m}}{dx}\right]\\
&+\left[\gamma^2x^2+A_{l m}-\frac{m^2}{1-x^2}\right]S_{l m}=0\, ,
\end{align}
with $\gamma=a\omega$. Its solutions are the spheroidal harmonics, $S_{l m}(x;\gamma)$, and $A_{l m}$ are the corresponding angular separation constants \cite{Berti:2005gp}. In the eikonal limit, an approximate expression for these constants is \cite{Yang:2012he}
\begin{equation}\label{Almapp}
A_{l m}(\gamma)\approx L^2-\frac{\gamma^2}{2}\left(1-\mu^2\right)\, ,
\end{equation}
with $L=l+1/2$. 
On the other hand, the radial equation is given by
\begin{align}\label{radial0}
&\Delta \frac{d}{dr}\left(\Delta \frac{d\psi_{lm}}{dr}\right)+V \psi_{lm}=0\, ,
\end{align}
where $V$ is an effective potential taking the form
\begin{equation}\label{Veff}
V=\left[\omega(r^2+a^2)-a m\right]^2-\lambda_{l m}\Delta\, ,
\end{equation}
and 
\begin{equation}
\lambda_{l m}\equiv A_{l m}(\gamma)-2m \gamma+\gamma^2\, .
\end{equation}

The QNMs correspond to the solutions of the radial equation \req{radial0} with outgoing boundary conditions both at the horizon and at infinity. In the eikonal limit, a solution can be found via the WKB approximation, which allows us to identify the QNM frequencies in terms of the minimum of the potential $r_0$ \cite{Iyer:1986np}. The real part $\omega_{R}$ is determined from the extremization conditions  
\begin{equation}\label{WKBR}
V\Big|_{r_0,\omega_{R}}=\frac{dV}{dr}\bigg|_{r_0,\omega_{R}}=0\, ,
\end{equation}
while the imaginary part is given by 
\begin{equation}\label{WKBI}
\omega_{I}=-\left(n+\frac{1}{2}\right)\frac{\Delta\sqrt{2\partial_{r}^2V}}{\partial_{\omega}V}\bigg|_{r_0, \omega_{R}}\, .
\end{equation}

A crucial observation is that, at extremality $a=M$, the horizon $r=M$ is always an extremum of the potential for the frequency $\omega_{R}=m \Omega$, where $\Omega=1/(2M)$ is the angular velocity of the horizon. However, it is not always a minimum, as required by the WKB approximation. By looking at the second derivative of the potential, we conclude that the horizon is a minimum only when 
\begin{equation}\label{Elmdef1}
\frac{1}{2}\frac{d^2V}{dr^2}\Big|_{r=M,a=M, \omega= m\Omega}=\frac{7}{4}m^2-A_{l m}(m/2)>0\, .
\end{equation}
Using the approximation \req{Almapp}, one can easily see that this only happens when $\mu$ is above a critical value $\mu> \bar\mu_{\rm cr}\approx 0.744$. As we show below, the corresponding modes have a vanishing imaginary part in the extremal limit and are therefore ZDMs. On the other hand, when $\mu<\bar\mu_{\rm cr}$ the horizon becomes a maximum, and instead, the potential develops a minimum outside the horizon. The corresponding modes acquire a non-zero imaginary part in the extremal limit and thus they are labeled as DMs. 
In addition, it was shown by \cite{Yang:2012pj,Yang:2013uba} that ZDMs also exist for $0\le \mu \le \bar\mu_{\rm cr}$ and they can be identified via matched asymptotic expansions --- the WKB method does not capture them since they are not related to a minimum of the potential. 

As a conclusion, the critical value $\bar\mu_{\rm cr}$ marks a ``phase boundary'' in the QNM spectrum: for $\mu> \bar\mu_{\rm cr}$ there are only ZDMs, while for $0\le \mu \le \bar\mu_{\rm cr}$ there are both ZDMs and DMs.  As noted in \cite{Yang:2012pj,Yang:2013uba}, in each level $(l, m)$ there are infinitely many ZDMs, but only around $l$ DMs. This branching of the spectrum implies a re-labeling of the overtones as extremality is approached. 

In this paper, we are especially interested in the behavior of modes that are close to the phase boundary $\mu\sim \bar\mu_{\rm cr}$, so we provide a few more details on those modes. For simplicity, we focus only on the modes that can be identified via the WKB approach, \textit{i.e.}, ZDMs for $\mu> \bar\mu_{\rm cr}$ and DMs for $\mu<\bar\mu_{\rm cr}$. 
We distinguish three different regimes that depend on the relative size of $\epsilon=1-a/M$ and $|\mu-\bar{\mu}_{\rm cr}|$ --- \textit{i.e.}, how close we are to extremality relative to how close we are to the critical value of $\mu$:
\begin{equation}\label{regimes}
\begin{aligned}
\text{Regime I: }& \quad \epsilon\ll |\mu-\bar{\mu}_{\rm cr}|^3\, , \quad \qquad \mu>\bar{\mu}_{\rm cr}\,,\\
\text{Regime II: }& \quad |\mu-\bar{\mu}_{\rm cr}|^3\ll \epsilon\ll 1\, ,\\
\text{Regime III: }& \quad \epsilon\ll |\mu-\bar{\mu}_{\rm cr}|^3\ll 1\, ,\quad \mu<\bar{\mu}_{\rm cr}\,.\\
\end{aligned}
\end{equation}
The first regime corresponds to the case in which the black hole is closer to extremality than $\mu$ is to the critical value, and $\mu$ can take any value above $\bar{\mu}_{\rm cr}$. In the second regime, we are closer to the critical point than to extremality, and it includes the case in which we are exactly at the critical point. Finally, the third regime corresponds to the case in which $\mu$ is below $\bar{\mu}_{\rm cr}$ but very close to it, and the black hole is arbitrarily close to extremality.\footnote{In reality,  quantum effects become important if $\epsilon$ is too small \cite{Iliesiu:2020qvm}. However, for macroscopic black holes the value of $\epsilon$ at which quantum effects become relevant is tiny, so the black hole remains classical across a big range of orders of magnitude of $\epsilon\ll 1$. We will thus neglect quantum effects.}
By solving the WKB equations \req{WKBR} and \req{WKBI} in the different regimes, we find the following expansions
\begin{align}\notag
\omega^{\rm I}=&m \Omega\left[1-\sqrt{2\epsilon} \sqrt{\frac{7}{4}-\frac{A_{l m}(m/2)}{m^2}}\right]\\\label{omegaIKerr}
&-\frac{i}{M}\left(n+\frac{1}{2}\right)\sqrt{\frac{\epsilon}{2}}+\mathcal{O}(\epsilon)\, ,\\\notag
\omega^{\rm II}=&m \Omega\left[1-\frac{3\epsilon^{2/3}}{2}\right]\\\label{omegaIIKerr}
&-\frac{i}{M}\left(n+\frac{1}{2}\right)\sqrt{\frac{3\epsilon}{4}}+\mathcal{O}(\epsilon)\, ,\\\notag
\omega^{\rm III}=&m \Omega\left[1+\frac{\eta^2(\mu-\bar\mu_{\rm cr})^2}{2}\right]\\\label{omegaIIIKerr}
&-i\left(n+\frac{1}{2}\right)\frac{\eta^{3/2}|\mu-\bar\mu_{\rm cr}|^{3/2}}{ 2M}+\ldots\, ,
\end{align}
where 
\begin{equation}
\eta=-\frac{1}{2}\frac{d}{d\mu}\left(\frac{A_{l m}(m/2)}{m^2}\right)\bigg|_{\bar{\mu}_{\rm cr}}\approx 2.33 \, .
\end{equation}
As we can see, in regimes I and II the imaginary part vanishes for $\epsilon\to 0$ and the frequency approaches the value $m\Omega$, although the dependence on $\epsilon$ is different in each case. In regime III, corresponding to DMs, the imaginary part is nonzero at $\epsilon=0$ (we are neglecting all $\mathcal{O}(\epsilon)$ terms in \req{omegaIIIKerr}), although it becomes arbitrarily small as $\mu\to \bar\mu_{\rm cr}$. Our goal in the rest of the paper is to understand how higher-derivative corrections modify these frequencies.

\section{Higher-derivative corrections to the QNM spectrum}\label{sec:QNMHD}
As a particular but relevant example, we focus our discussion on an eight-derivative extension of GR given by the quartic action
\begin{equation}\label{eq:ISO}
S=\frac{1}{16\pi G}\int d^{4}x\sqrt{|g|}\left[R+\alpha\left(\mathcal{R}_{2}^2+ \tilde{\mathcal{R}}_{2}^2\right)\right]\, ,
\end{equation}
where $\mathcal{R}_{2}=R_{\mu \nu \rho \sigma} R^{\mu \nu \rho \sigma}$,  $\tilde{\mathcal{R}}_{2}=R_{\mu \nu \rho \sigma} \tilde{R}^{\mu \nu \rho \sigma}$ and $\tilde{R}^{\mu \nu \rho \sigma}=\frac{1}{2} \epsilon^{\mu \nu \alpha \beta} R_{\alpha \beta}{ }^{\rho \sigma}$ is the dual Riemann tensor.  The coupling constant $\alpha$ has dimensions of [length]$^6$ and its order of magnitude is related to a new length scale $\ell_{\rm new}$, so that $\alpha\sim \ell_{\rm new}^{6}$. 
For a black hole solution of mass $M$, we introduce the dimensionless coupling
\begin{equation}
\hat\alpha=\frac{\alpha}{M^6}\, ,
\end{equation}
that controls the size of the corrections. We restrict to the regime $|\hat\alpha|\ll 1$.

It was shown by \cite{Cano:2024wzo} that this theory is the unique EFT extension of GR, up to eight derivatives, that leads to non-birefringent graviton propagation. As a consequence, it also preserves the isospectrality of QNMs in the eikonal limit. Taking advantage of these properties, Ref.~\cite{Cano:2025mht} has been able to perform a detailed analysis of eikonal QNMs of black holes with arbitrary rotation, and this is our main reason to focus on this theory. On the other hand, \req{eq:ISO} is also interesting from a top-down point of view as it coincides with the quartic effective action of type IIB string theory \cite{Gross:1986iv}. 

As shown by \cite{Cano:2025mht}, gravitational perturbations of large momentum in the theory \req{eq:ISO} are governed by a master scalar equation
 \begin{equation}\label{eq:wave_eq1}
\nabla^2\Phi+64\alpha\tensor{R}{^{\mu}_\alpha^\nu_\beta}\tensor{R}{^{\rho\alpha\sigma\beta}}\nabla_{\mu}\nabla_{\nu}\nabla_{\rho}\nabla_{\sigma}\Phi=0\, ,
\end{equation}
where the Riemann tensor is the one of the background spacetime. Crucially, this equation captures perturbations of arbitrary parity, which is a consequence of isospectrality. Another interesting aspect of \eqref{eq:wave_eq1} is that, in order to investigate the corrections to the Kerr eikonal QNMs, we can neglect the corrections to the Kerr metric and simply consider \eqref{eq:wave_eq1} on the Kerr geometry. This is because the corrections to the background are subleading in the eikonal limit with respect to the second term in \eqref{eq:wave_eq1}. This is advantageous as corrections to the highly-spinning Kerr metric are notoriously difficult to obtain \cite{Cano:2019ore,Lam:2025elw,Lam:2025fzi}.

When evaluated on the Kerr background, the scalar field in equation \eqref{eq:wave_eq1} can be decomposed into spheroidal harmonics as\footnote{This kind of decomposition and the discussion below would also apply if we included the corrections to the Kerr metric \cite{Cano:2020cao,Ghosh:2023etd}.} 
$\Phi=\sum_{l', m} e^{-i\omega t+i m\phi} S_{l' m}(x; a \omega) \psi_{l' m}(r)\, .$
This series contains a leading harmonic, say $l'=l$, which is the only one that survives in the GR limit $\alpha\to 0$; all the other harmonics are of order $\alpha$. Projecting the equation onto the leading harmonic $S_{l m}$ leads to a modified radial Teukolsky equation at first order in $\alpha$, 
\begin{equation}\label{radial}
\Delta \frac{d}{dr}\left(\Delta \frac{d\psi_{lm}}{dr}\right)+V_{\alpha} \psi_{lm}=0\, ,
\end{equation}
where the potential $V_{\alpha}$ reads
\begin{equation}\label{Veffalpha}
V_{\alpha}=\left[\omega(r^2+a^2)-a m\right]^2-\Delta\left(\lambda_{l m}+\hat{\alpha} U_{l m}\right)\, .
\end{equation}
Naturally, this reduces to the usual Teukolsky potential \req{Veff} when $\alpha=0$. 
The correction to the potential $U_{l m}$ is given explicitly by the integral

\begin{widetext}
\begin{equation}
U_{l m}=-1152\lambda_{l m}^2 \left(\frac{M}{r}\right)^{8}\frac{1}{K(-q)}\int_{0}^{\pi/2} \frac{du}{\left(1+\frac{a^2x_0^2}{r^2}\sin^2 u\right)^{4}\sqrt{1+q \sin^2 u}}\,,
\end{equation}
\end{widetext}
where
\begin{align}\label{x0}
x_0^2&=\frac{-A_{l m}+\gamma^2+\sqrt{(A_{l m}+\gamma^2)^2-4m^2 \gamma^2}}{2\gamma^2}\, ,\\
\label{qsol}
q&=\frac{x_0^2\gamma^2}{A_{l m}-(1-x_0^2)\gamma^2}\, ,
\end{align}
and $K(-q)$ is the elliptic integral of the first kind. With these expressions at hand, one can then apply the WKB formulas \req{WKBR} and \req{WKBI} to find the QNM frequencies. We consider two different approaches: a perturbative expansion in $\hat\alpha$, and an exact treatment of the frequencies.

\subsection{Linear corrections to the QNMs}
Since we are assuming $|\hat\alpha|\ll 1$, it is natural to assume that we can expand the modification of the frequencies as a perturbative series in $\hat\alpha$. Thus, we write
\begin{equation}
\omega=\omega^{\rm Kerr}+\hat\alpha \delta\omega+\mathcal{O}(\hat\alpha^2)\, ,
\end{equation}
where $\omega^{\rm Kerr}$ is the corresponding QNM frequency for Kerr black holes in GR, and $\delta\omega$ is the coefficient of the linear-in-$\alpha$ correction. General formulas for $\delta\omega$ were provided by \cite{Cano:2025mht}, so we refer to that work for more details on this computation.

We are interested in the behavior of $\delta\omega$ in the three regimes defined in \req{regimes}.  
Within regime I, we find that the corrections to the real and imaginary parts of the QNM frequencies behave as $\delta\omega_{R}=\mathcal{O}\left(\epsilon^{1/2}\right)$ and $\delta\omega_{I}=\mathcal{O}(\epsilon)$ when $\epsilon\to 0$. Thus, the approach to extremality is as in GR and corrections do not introduce any qualitative difference. 

\begin{figure}[t!]
  	\centering
  	\includegraphics[width=0.48\textwidth]{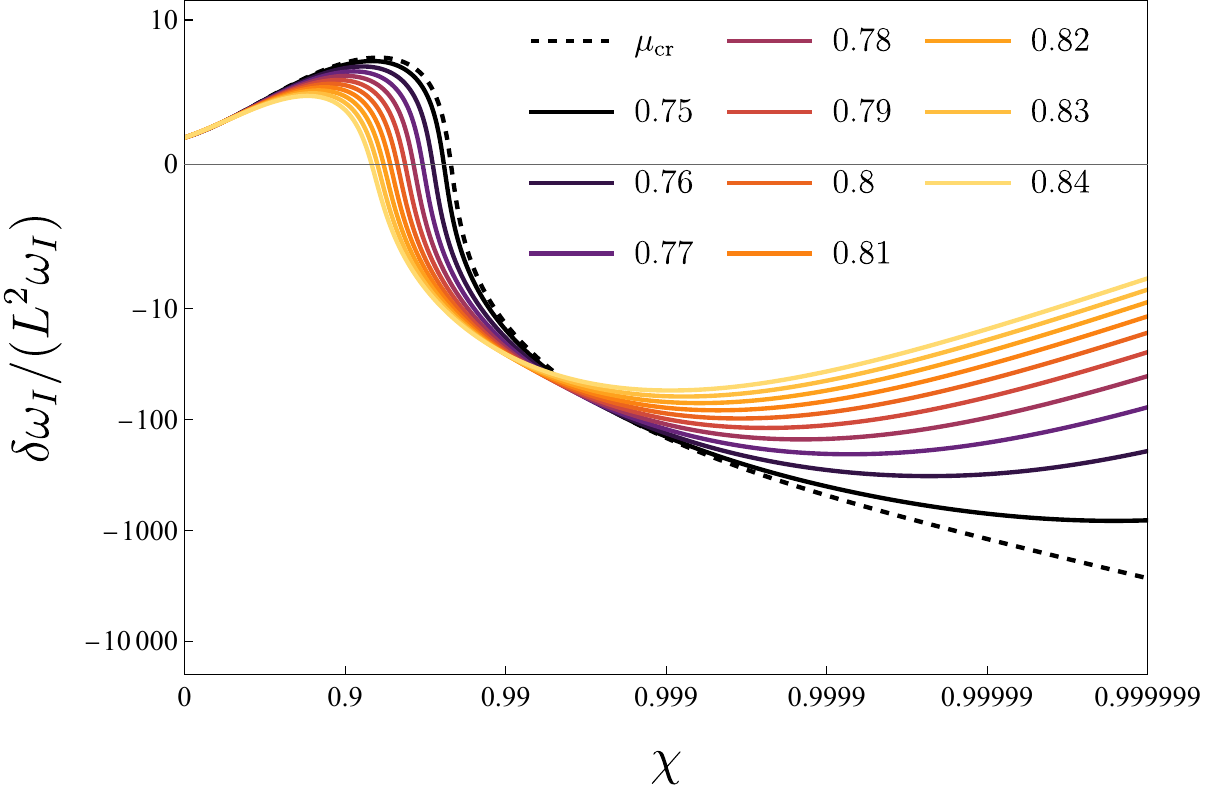}
  	\caption{Relative corrections to the imaginary part of the QNM frequencies as a function of the dimensionless spin $\chi=a/M$ for several values $\mu\ge \bar\mu_{\rm cr}$.  The corrections become arbitrarily large when $\mu$ approaches $\bar\mu_{\rm cr}$ and we take $\chi\to 1$ (observe that the plot is in a log-log scale). For $\mu= \bar\mu_{\rm cr}$ (dashed line), the relative corrections diverge at extremality --- see \req{relII}.}
  	\label{fig:wI}
  \end{figure}

Things are more interesting as we get close to the critical point. In order to write the results, we introduce the coefficient\footnote{This is related to the quantity $C(\bar{\mu}_{\rm cr})$ introduced in  \cite{Cano:2025mht} by $\xi=-1152 C(\bar{\mu}_{\rm cr})$. } 
\begin{equation}
\xi \equiv \frac{U_{l m}}{m^4}\Big|_{a=M,\, r=M,\, \omega=L \bar\mu_{\rm cr}\Omega,\, m=L\bar\mu_{\rm cr}}\approx \, -601.27\,.
\end{equation}
In regime II, we find
\begin{align} 
	\delta\omega_{R}&=\frac{L^3\bar{\mu}_{\rm cr}^3}{4M}\xi \epsilon^{1/3}+\ldots \,,
	\\\label{domegaI2}
		\delta\omega_{I}&=- \left(n+\frac{1}{2}\right)\frac{L^2\bar{\mu}_{\rm cr}^2}{8\sqrt{3}M} \xi \epsilon^{1/6}+\ldots \,,
\end{align}
and now something remarkable happens; while overall the imaginary part tends to zero as $\epsilon\to 0$ --- meaning that the modes become infinitely long-lived --- the corrections go to zero \emph{slower} than the GR contribution \req{omegaIIKerr}. This implies that, for black holes close enough to extremality, the corrections may overcome the GR result, leading to a large change in the modes' damping time. In fact, the relative corrections to GR diverge in the extremal limit, 
\begin{equation}\label{relII}
\frac{\delta\omega_{I}}{\omega_{I}^{\rm Kerr}}\approx \frac{L^2\bar{\mu}_{\rm cr}^2\xi}{12}\epsilon^{-1/3}.  
\end{equation}\
We illustrate this in Figure~\ref{fig:wI}, where we show the relative correction to the imaginary part as a function of the black hole spin, for several values of $\mu\ge \bar{\mu}_{\rm cr}$. 

On the other hand, in regime III we get

\begin{align} 
	\delta\omega_{R}&= \frac{L^3\bar{\mu}_{\rm cr}^3}{4M}  \xi \eta |\mu-\bar{\mu}_{\rm cr}| +\dots  \,,
	\\\label{domegaI3}
		\delta\omega_{I}&= -\left(n+\frac{1}{2}\right)\frac{3}{8 M}L^2\bar{\mu}_{\rm cr}^{2} \xi\eta^{1/2} |\mu-\bar{\mu}_{\rm cr}|^{1/2}   + \ldots \,.
\end{align}
We remark that here we are dropping all $\mathcal{O}(\epsilon)$ terms, so these formulas represent the DMs arbitrarily close to extremality when $\mu\to \bar{\mu}_{\rm cr}$.  Similarly to regime II, we observe that the correction to the imaginary part goes to zero slower than the GR value \req{omegaIIIKerr} when $\mu\to \bar{\mu}_{\rm cr}$. This leads again to arbitrarily large corrections to the damping times of modes close to the critical value and to a blowup of the relative corrections
\begin{equation}\label{relIII}
\frac{\delta\omega_{I}}{\omega_{I}^{\rm Kerr}}\approx \frac{3L^2\bar{\mu}_{\rm cr}^2\xi}{4\eta  |\mu-\bar{\mu}_{\rm cr}|}.  
\end{equation}\

A natural question at this point is whether these effects are physically meaningful and represent an actual amplification of new physics, or if they are pointing out a breakdown of the EFT. To answer this question, we re-examine these QNMs without performing an expansion in $\hat\alpha$. 

\subsection{QNMs near extremality: exact analysis}
We can obtain analytically the QNMs of the modified Teukolsky equation \req{radial} near extremality without resorting to an expansion in small $\hat\alpha$. The key observation is that, when evaluated at $\omega=m \Omega$ and at extremality, the potential reads
\begin{equation}
\begin{aligned}
V_{\alpha}\Big|_{a=M, \omega= m\Omega}=&(r-M)^2\bigg[\frac{m^2}{4}\left(\frac{r}{M}+1\right)^2\\
&+\frac{3}{4}m^2-A_{l m}-\hat{\alpha} U_{l m}\bigg]\, .
\end{aligned}
\end{equation}
Thus, the horizon $r=M$ is always a stationary point of the potential, like in the Kerr case. However, it is only the relevant point for the WKB method if it is a minimum. Therefore, we have to look at the sign of the second derivative of $V_{\alpha}$, 
\begin{equation}\label{Elmdef}
\frac{1}{2}\frac{d^2V_{\alpha}}{dr^2}\Big|_{r=M,a=M, \omega= m\Omega}=\frac{7}{4}m^2-A_{l m}-\hat{\alpha} U_{l m}\equiv E_{l m}\, .
\end{equation}
When $E_{l m}>0$, the horizon is a minimum of $V_{\alpha}$ and from the WKB method we obtain ZDMs that tend to $\omega\to m \Omega$ in the extremal limit. We can therefore perform an expansion of the WKB formulas \req{WKBR} and \req{WKBI} around $\omega=m \Omega$, $\epsilon=0$ and $r=M$ to find the frequencies. 
When $E_{l m}<0$, the potential develops instead a minimum outside the horizon and we get DMs. If $E_{l m}$ is small in absolute value, the minimum lies close to $r=M$ and the frequency is close to $\omega = m \Omega$, so we can again find the frequencies through an expansion. In this way, we can find analytic expressions for all the three regimes in \req{regimes}, with the difference that now $E_{l m}/L^2$ plays the role of $\mu-\bar{\mu}_{\rm cr}$. 

We obtain the following results for the full QNM frequencies, without performing any expansion in $\hat\alpha$, 
\begin{align}
\omega^{\rm I}=&m \Omega-\frac{\sqrt{2\epsilon E_{l m}}}{2M}-\frac{i}{M}(n+1/2)\sqrt{\frac{\epsilon}{2}}+\mathcal{O}(\epsilon)\, ,\\\notag
\omega^{\rm II}=&m \Omega-\frac{3m\epsilon^{2/3}\left(1-M \hat\alpha U'_{l m}/m^2\right)^{1/3}}{4M}\\
&-\frac{i}{M}(n+1/2)\sqrt{\frac{3\epsilon}{4}}+\mathcal{O}(\epsilon)\, ,\\\label{omegaIII}
\omega^{\rm III}=&m \Omega+\frac{E_{l m}^2- i(n+1/2)|2E_{l m}|^{3/2}}{16 M m (m^2-M \hat\alpha U'_{l m})}+\ldots\, ,
\end{align}
where $U_{lm}'=dU_{lm}/dr\big|_{r=a=M,\, \omega=L \bar\mu_{\rm cr}\Omega,\, m=L\bar\mu_{\rm cr}}$. 
As we can see, these expressions are perfectly finite and they very much resemble the behavior of the Kerr QNMs in equations~\req{omegaIKerr}, \req{omegaIIKerr} and \req{omegaIIIKerr}.  
However, the most important difference is that the critical point, given by the condition $E_{l m}=0$, is modified. In fact, if we evaluate $E_{l m}$ for $\mu$ close to $\bar{\mu}_{\rm cr}$ --- the original critical point in GR --- we obtain
\begin{equation}\label{Elmexp}
\frac{E_{l m}}{m^2}\approx 2\eta (\mu-\bar{\mu}_{\rm cr})-\hat\alpha L^2\bar{\mu}_{\rm cr}^2\xi\, .
\end{equation}
Thus, solving for $E_{lm}=0$, we find that the actual critical point is located at 
\begin{equation}\label{mucrcorrection}
\mu_{\rm cr}\approx \bar{\mu}_{\rm cr}+ \hat\alpha\delta \mu_{\rm cr}\, ,\quad \delta\mu_{\rm cr}=\frac{L^2\bar{\mu}_{\rm cr}^2}{2\eta}\xi\, .
\end{equation}
The expansion of this additional dependence on $\alpha$ is the reason behind the divergences in \req{relII} and \req{relIII}. This is most easily seen in regime III. If we assume $|\mu-\mu_{\rm cr}|\ll1$ and we use \req{omegaIII} and \req{Elmexp}, we see that the imaginary part in regime III is approximately
\begin{equation}\label{omegaIfull}
\begin{aligned}
\frac{\omega_{I}}{(n+1/2)}\approx &-\frac{\eta^{3/2}|\mu-\mu_{\rm cr}|^{3/2}}{2M}\left(1+\mathcal{O}(\hat\alpha)\right)\, .
\end{aligned}
\end{equation}
If we then expand this expression in $\hat\alpha$ using \req{mucrcorrection}, and keep only the leading terms in $|\mu-\bar\mu_{\rm cr}|$, we obtain 
\begin{equation}\label{omegaIexp}
\begin{aligned}
\frac{\omega_{I}}{(n+1/2)}\approx& -\frac{\eta^{3/2}}{2M}|\mu-\bar\mu_{\rm cr}|^{3/2}\\
&-\frac{3\eta^{3/2}}{4M}\hat\alpha  \delta\mu_{\rm cr}|\mu-\bar{\mu}_{\rm cr}|^{1/2}+\ldots\, ,
\end{aligned}
\end{equation}
and we exactly reproduce the Kerr result \req{omegaIIIKerr} and the linear correction \req{domegaI3}. This is telling us that the fact that the linear in $\hat\alpha$ corrections seem to overcome the GR contribution is an effect of the critical point being modified. This new computation shows that the linear correction \req{domegaI3} becomes inaccurate when $\mu-\bar\mu_{\rm cr}$ is of the same order as $\hat\alpha\delta\mu_{\rm cr}$. In that regime, one should use instead the exact expression \req{omegaIII} or \req{omegaIfull}. 
For similar reasons, the change of the critical point also explains why the relative corrections $\delta\omega_{I}/\omega_{I}^{\rm Kerr}$ in regime II [equation \req{relII}] seem to blow up when $\epsilon\to 0$. 

The blow-up of the relative corrections is therefore signaling a large change in the spectrum, since a modification in the critical point means a change in the nature of the modes close to it.  If $\hat\alpha\delta\mu_{\rm cr}>0$, the modes in the range $0<\mu-\bar\mu_{\rm cr}<\hat\alpha\delta\mu_{\rm cr}$ are ZDMs in GR but become DMs in the higher-derivative theory. Conversely, if $\hat\alpha\delta\mu_{\rm cr}<0$, the modes with $0>\mu-\bar\mu_{\rm cr}>\hat\alpha\delta\mu_{\rm cr}$ are DMs in GR and turn into ZDMs when the corrections are included. This is a large change in the spectrum, since ZDMs and DMs have very different lifetimes near extremality. 
However,  this discussion only applies to the modes identified via the WKB method. Analogously to GR, we expect that ZDMs exist in the full range $\mu\ge 0$, but the WKB method only captures those above $\mu_{\rm cr}$. Therefore, what is really happening is that the number of DMs is affected by the corrections: $\hat\alpha\delta\mu_{\rm cr}>0$ will potentially increase the number of DMs while $\hat\alpha\delta\mu_{\rm cr}<0$ will decrease it. 
In any case, modes in the critical band
\begin{equation}\label{critband}
|\mu-\bar\mu_{\rm cr}|<|\hat\alpha\delta\mu_{\rm cr}|
\end{equation}
receive large corrections regardless whether the corrections push them to the other side of the phase boundary or away from it. 

\section{Regime of validity} \label{sec:regimeofvalidity}
Our previous discussion has shown that the growth of the relative corrections to the imaginary part of the QNMs near the critical point is a genuine effect and it is not related, in principle, to a breakdown of the EFT approach. Moreover, this effect is clearly general, as we can expect it to appear in any other theory beyond GR that modifies the Teukolsky equation. 
However, we still have to answer whether large modifications of the spectrum --- \textit{i.e.} a change in the number of DMs --- can \emph{actually} take place in our current setup while remaining in a regime in which the theory \req{eq:ISO} is sensible. 

\subsection{Wilsonian point of view}
We consider two different notions for a theory being ``sensible''. The first one is the usual top-down Wilsonian approach in which we assume that the higher-derivative terms are generated by integrating out massive degrees of freedom, whose masses $E_{\rm UV}$ fix the scale of the coupling constants. In our setup, this means that the scale of the coupling $\alpha\sim \ell_{\rm new}^6$ is given by the UV scale, $\ell_{\rm new}\sim \ell_{\rm UV}=E_{\rm UV}^{-1}$. Then, the Wilsonian approach would tell us that we can only trust the predictions of the EFT as long as distances larger than $\ell_{\rm UV}$ are involved because at that length scale one would start seeing the massive degrees of freedom. This implies, in particular, that the EFT should only be applied to black holes of radius larger than $\ell_{\rm UV}$ and that GWs should have wavelengths larger than $\ell_{\rm UV}$.
Since the radius of an extremal black hole is $M$ and the wavelength of a wave with harmonic number $l$ is roughly $M/l$, the EFT is valid as long as 
\begin{equation}\label{EFTregime1}
M\gg \ell_{\rm UV}\, ,\quad l\ll \frac{M}{\ell_{\rm UV}}\, .
\end{equation}
Let us then determine whether there can be large corrections to the spectrum while remaining consistent with these conditions. As we saw earlier, the modes affected are those in the critical band \req{critband}. Using \req{mucrcorrection} and \req{EFTregime1}, we find that $\mu$ should satisfy
\begin{equation}\label{critband2}
|\mu-\bar\mu_{\rm cr}|<|\hat\alpha|\frac{L^2\bar{\mu}_{\rm cr}^2}{2\eta}|\xi| \ll \frac{71.4}{L^4} \left(\frac{\ell_{\rm new}}{\ell_{\rm UV}}\right)^6\, ,
\end{equation}
where we have introduced the numerical values of the different constants. 
Thus, up to a numerical factor,  $\mu$ must lie within a distance $1/L^4$ away from $\bar\mu_{\rm cr}$. However, we must recall that  $\mu$ is a rational number given by \req{mudef}, while the critical value $\bar\mu_{\rm cr}$ is, in all likelihood, an irrational number. Dirichlet's theorem states that, in general, the best approximation to an irrational number with rational numbers with denominator\footnote{Since $L$ is semi-integer, $\mu$ can be written as a fraction with denominator $2L$.} $L$ is $|\mu-\bar\mu_{\rm cr}|\lesssim 1/L^2$. The bound \req{critband2} is smaller than this for large $L$, so that, in general, there are no modes in the critical band. Thus, we have achieved the quite interesting conclusion that, for large $L$, the number of DMs cannot change within the standard regime of validity of EFT, and that corrections cannot reach order one.  
Still, the modes near $\bar\mu_{\rm cr}$ do receive larger corrections than other modes. In fact, if we apply for instance \req{relIII} and we use that $|\mu-\bar\mu_{\rm cr}|\lesssim 1/L^2$, we get that the maximum size of the corrections is
\begin{equation}
\frac{\hat\alpha\delta\omega_{I}}{\omega_{I}^{\rm Kerr}}\sim \hat\alpha L^4 \lesssim |\hat\alpha|^{1/3}\, ,
\end{equation}
where we used that the maximum value of $L$ is $L\lesssim M/\ell_{\rm UV}\sim |\hat\alpha|^{-1/6}$. Since $|\hat\alpha|\ll 1$, this implies that the corrections are much larger than $\mathcal{O}(\hat\alpha)$.

On the other hand, these arguments only apply for $L\to \infty$ since they rely on the scaling of different quantities. There could exist low-$l$ modes in the critical band as long as $\ell_{\rm UV}$ is just slightly smaller than $\ell_{\rm new}$. In fact,  if we again assume that $|\mu-\bar{\mu}_{\rm cr}|\sim 1/L^2$, we see that \req{critband2} is satisfied for $L^2\ll 71.4(\ell_{\rm new}/\ell_{\rm UV})^6$. Thus, if $\ell_{\rm new}\gtrsim \ell_{\rm UV}$, the term on the right-hand side is large enough so that low values of $L$ satisfy this criterion.  

\subsection{Self-consistency of the effective action}
It is possible to consider a different notion of a ``sensible theory''. Namely, from a bottom-up perspective, we can say that the theory remains sensible as long as the higher-derivative expansion is convergent, \textit{i.e}, terms with more derivatives remain subleading with respect to terms with less derivatives. In this approach, we remain agnostic about the hypothetical UV completion of the effective action, so we do not assume that the scale of $\alpha\sim \ell_{\rm new}^6$ is necessarily the same as the UV scale.  To determine the conditions in which the theory remains sensible, we investigate the effect of extra higher-derivative corrections to both the background solution and to gravitational waves. 

One can see that, in order for the higher-derivative expansion of the background spacetime to be convergent, the mass (equivalently the radius) must satisfy $M\gg \ell_{\rm new}$, just like before. Equivalently, this means that $|\hat\alpha|\ll 1$. 

In the case of gravitational perturbations, the question that we would like to answer is whether there are modes in the critical band \req{critband} for which the higher-derivative expansion is still under control. As we noted before, since $\mu$ is a rational number, we can find many modes for which $|\mu-\bar\mu_{\rm cr}|\lesssim 1/L^2$. Therefore, in order for the inequality \req{critband} to be satisfied, we need to consider values of $L$ such that\footnote{Actually, if we keep the numerical coefficients, a more accurate bound is $71.4 L^4\left(\frac{\ell_{\rm new}}{M}\right)^6\gtrsim 1$, which is less strict than \req{critcond}. However, the factors are irrelevant for our argument in this section, so we remove them for simplicity. }
\begin{equation}\label{critcond}
L^4\left(\frac{\ell_{\rm new}}{M}\right)^6\gtrsim 1\, .
\end{equation}
We ask whether in this regime the higher-derivative expansion of gravitational perturbations is under control.  A way to answer this question is to analyze the effect of additional higher-derivative corrections to the dispersion relation of large-momentum gravitational perturbations. The dispersion relation in fact controls the dynamics in the eikonal limit and is essentially equivalent to the master equation \req{eq:wave_eq1} --- see \cite{Cano:2025mht}. 

For our theory \req{eq:ISO}, the dispersion relation takes the schematic form
\begin{equation}\label{disp1}
k^2+\ell_{\rm new}^6 k^4 R^2=0\, ,
\end{equation}
where $k$ represents the momentum and $R$ the background curvature tensor. Let us then include additional higher-derivative corrections and investigate their size with respect to the terms already present in \req{disp1}.  For simplicity, we first assume that the corrections also preserve the eikonal-isospectral condition, so that they are independent of the polarization. 

For large momentum, the leading contribution of any higher-derivative term to the dispersion relation consists of $2n$ momenta contracted with a background tensor formed from the curvature tensor and its derivatives.  Therefore, a generic contribution can be written as
\begin{equation}
k^2+\ell_{\rm new}^6 k^4 R^2+\ell_{\rm new}^{2(n+p+q-1)}k^{2n}\nabla^{2p}R^{q}=0\, ,
\end{equation}
where the covariant derivatives act on the Riemann tensors and the indices can in principle be contracted in many different forms.  We note that we are assuming that the additional terms also enter at the scale $\ell_{\rm new}$. 

Let us then evaluate the relative magnitude of each term in this equation. The momentum is $|k|\sim L/M$ while near the horizon we have $|\nabla|\sim 1/M$ and $|R|\sim 1/M^2$. 
The ratio between the magnitudes in the second and first term is
\begin{equation}
\frac{\ell_{\rm new}^6 |k|^4 |R|^2}{|k|^2}\sim L^2 \left(\frac{\ell_{\rm new}}{M}\right)^{6}\, ,
\end{equation}
while the ratio between the third and the second is
\begin{equation}
\frac{\ell_{\rm new}^{2(n+p+q-1)}|k|^{2n}|\nabla^{2p}R^{q}|}{\ell_{\rm new}^6 |k|^4 |R|^2}\sim L^{2n-4}\left(\frac{\ell_{\rm new}}{M}\right)^{2(n+p+q-4)}\, .
\end{equation}
Now, let us assume that the condition \req{critcond} is satisfied, while at the same time $\ell_{\rm new}/M\ll 1$.  In this regime the first ratio becomes
\begin{equation}
\frac{\ell_{\rm new}^6 |k|^4 |R|^2}{|k|^2}\sim \left(\frac{\ell_{\rm new}}{M}\right)^{3}\ll 1\, ,
\end{equation}
meaning that the size of the eight-derivative corrections is still small compared to GR.  This is already an indication that the higher-derivative expansion is under control. But now the crucial question is: could the higher-order terms become larger than the eight-derivative terms? This would actually imply the breakdown of the derivative expansion. 
In the regime \req{critcond}, we get
\begin{equation}\label{ratioexp}
\frac{\ell_{\rm new}^{2(n+p+q-1)}|k|^{2n}|\nabla^{2p}R^{q}|}{\ell_{\rm new}^6 |k|^4 |R|^2}\sim \left(\frac{\ell_{\rm new}}{M}\right)^{2(p+q-n/2-1)}\, ,
\end{equation}
so the problem boils down to determining if the exponent $p+q-n/2-1$ can be negative. 
Let us observe the following: since $k_{\mu}$ is null in GR, contractions of two momenta yield $k_{\mu}k^{\mu}= \mathcal{O}\left(\alpha\right)$, which is subleading and we therefore neglect it. On the other hand, a Riemann tensor can at most be contracted with two momenta --- a contraction with three of four is identically zero. Therefore, the way to maximize the number of momenta in the dispersion relation is to contract each $\nabla$ with one momentum and each Riemann tensor with two momenta. Since there are $2p$ nablas and $q$ Riemann tensors, we conclude that the maximum number of momenta that yields a non-zero contribution is
\begin{equation}\label{nbound}
n\le p+q\, .
\end{equation}
On the other hand, we are considering terms with more than eight derivatives, which implies
\begin{equation}
2(n+p+q)>8\, .
\end{equation}
Combining these two inequalities, we also get
\begin{equation}
4<n+p+q\le  2(p+q) \Rightarrow p+q>2\, .
\end{equation}
Therefore, the exponent in \req{ratioexp} satisfies
\begin{equation}
p+q-\frac{n}{2}-1\ge p+q-\frac{p+q}{2}-1=\frac{p+q}{2}-1>0\, ,
\end{equation}
and it must be strictly positive. Hence, the ratio \req{ratioexp} is much smaller than one. 

Finally, we can easily generalize this argument to the case in which the additional corrections break the isospectral property. In this case, the higher-derivative terms can also depend on the polarization tensor $e_{\mu\nu}$. However, since this tensor is transverse to the momentum, this means that $e_{\mu\nu}$ must necessarily be contracted either with the curvature tensor, with the covariant derivative or with itself, and therefore, it always reduces the number of free indices to which $k_{\mu}$ can be contracted. Hence, the number of momenta in a polarization-dependent term in the dispersion relation also satisfies the bound \req{nbound} and consequently, the ratio \req{ratioexp} also remains very small for those theories.  

This argument, which is completely general, implies that any additional higher-derivative corrections to the dispersion relation are subleading with respect to the original quartic term, as long as they appear at the same scale. Therefore, the effect of the quartic term will not be significantly affected by higher-order corrections. In this sense, we conclude that the phenomenon of modes crossing the phase boundary can actually happen in a regime in which the theory, in a classical sense, is well-behaved.

\section{Sensitive modes for low $l$} \label{sec:sensitivemodes}
Our previous arguments apply only to the eikonal limit. However, since the spectrum of modes is discrete, it can happen that some modes lie exceptionally close to the critical line separating DMs and ZDMs, and so they will be especially sensitive to corrections. In addition, lower $l$ modes are the ones with astrophysical interest. 

Although we cannot obtain QNMs of highly-spinning black holes for low $l$ with our current methods, we can however determine which modes are more sensitive and give an estimation of the effect of the corrections. The key to achieve this is to analyze the condition that determines the existence or absence of DMs. 
As explained in \cite{Yang:2013uba}, the idea is again to determine whether the effective potential of the Teukolsky equation develops a minimum outside the horizon in the extremal limit. Establishing this fact is not straightforward, since the Teukolsky equation for gravitational perturbations is complex  (except in the eikonal limit, where it coincides with the equation for a scalar field perturbations), so the notion of a minimum is not well defined.  Instead, one has to transform the Teukolsky equation into a real equation, as done by \cite{Detweiler:1977gy}, and then analyze the extrema of the corresponding potential \cite{Yang:2013uba}. The result of this analysis for Kerr black holes is that the potential does not have a minimum outside the horizon --- and therefore, there are no DMs --- when the following condition holds
\begin{equation}\label{noDM1}
{}_{s}E_{l m}^{\rm Kerr}=\frac{7m^2}{4}-s(s+1)- {_{s}A}_{l m}(m/2)>0\, .
\end{equation}
Here $s$ is the spin of the perturbation ($s=\pm 2$ for gravitational perturbations) and ${_{s}A}_{l m}$ are the separation constants of the spin-weighted spheroidal harmonics. We note that this quantity is essentially the same as \req{Elmdef} up to spin-dependent corrections, and it simply corresponds to the second derivative of the potential at the horizon, so that its sign determines if the horizon is a maximum or a minimum.  

Now, to find the condition \req{noDM1}, we do not really need to use the full Teukolsky equation; it is enough to know the Teukolsky equation in the near-horizon region of extremal black holes, for frequencies around $\omega= m\Omega$. Fortunately, the modification of such near-horizon Teukolsky equation due to higher-derivative corrections is known, and was determined in \cite{Cano:2024bhh}. The modification of the radial equation is very simple, as it only entails a change in the angular separation constants, ${_{s}A}_{l m}\to {_{s}A}_{l m}+\hat\alpha \delta A_{l m}^{\pm}$. The correction $\delta A_{l m}^{\pm}$ is the same for both $s=+2$ and $s=-2$, but it depends on the parity of the perturbation, denoted by the superscript $\pm$.  For our eikonal-isospectral theory \req{eq:ISO} we still have $\delta A_{l m}^{+}\neq \delta A_{l m}^{-}$ for finite $l$, but the values converge to each other as $l$ increases. We show the values of $\delta A_{l m}^{\pm}$ up to $l=4$ in Table~\ref{deltaAtab}.

\begin{table}[]
	
	\sisetup{
		table-format = 2.2,   
		detect-all,
		table-number-alignment = center
	}
	
	\centering
	\def\arraystretch{1.4}
	\setlength{\tabcolsep}{2pt}
	\begin{tabular}{
			|>{\centering\arraybackslash}p{0.6cm}||
			S[table-format=3.2, table-column-width=2.8em]|
			S[table-format=3.2, table-column-width=2.8em]|
			S[table-format=3.2, table-column-width=3.0em]|
			S[table-format=3.2, table-column-width=3.0em]|
			S[table-format=3.2, table-column-width=3.2em]|
			S[table-format=3.2, table-column-width=3.2em]|
		}
		\hline
		\diagbox[dir=NW,width=2.2em,height=2.4em,innerleftsep=1pt,innerrightsep=3pt]{$m$}{$l$}
		& \multicolumn{2}{c|}{2}
		& \multicolumn{2}{c|}{3}
		& \multicolumn{2}{c|}{4} \\ \hline
		& $\delta A^{+}$ & $\delta A^{-}$
		& $\delta A^{+}$ & $\delta A^{-}$
		& $\delta A^{+}$ & $\delta A^{-}$ \\ \hline\hline
		0 & -8.05 & -0.88 & -8.38 & -10.16 & -24.68 & -19.58 \\ \hline
		1 & -1.23 & -4.91 & -12.91 & -6.07 & -20.52 & -21.65 \\ \hline
		2 & -1.67 & -1.26 & -4.61 & -10.55 & -22.13 & -15.85 \\ \hline
		3 &  &  & -4.60 & -3.67 & -11.07 & -18.75 \\ \hline
		4 &  &  &  &  & -9.61 & -8.10 \\ \hline
	\end{tabular}
	\caption{Values of $\delta A_{lm}^{\pm}$ for $l=2,3,4$. Each value should be multiplied by $10^4$.}
	\label{deltaAtab}
\end{table}

Taking this into account, we can immediately write down the condition for the horizon to be a minimum of the potential by generalizing \req{noDM1}:
\begin{equation}\label{noDM2}
{}_{s}E_{l m}=\frac{7m^2}{4}-s(s+1)- {_{s}A}_{l m}-\hat\alpha \delta A_{l m}^{\pm}>0\, ,
\end{equation}
where both separation constants are evaluated at $a\omega=m/2$. For modes for which ${}_{s}E_{l m}^{\rm Kerr}$ was already very close to zero, the effect of the corrections can become important even if $\hat\alpha\ll 1$. In order to search for the most sensitive modes, we consider the ratio\footnote{This is the same for $s=+2$ and $s=-2$ due to the property ${_{-s}A}_{l m}={_{s}A}_{l m}+2s$ of the angular separation constants. }
\begin{equation}
\Delta_{l m}^{\pm}=\frac{\delta A_{l m}^{\pm}}{\frac{7m^2}{4}-s(s+1)- {_{s}A}_{l m}}\, ,
\end{equation}
which represents the relative correction (without $\hat\alpha$) to ${}_{s}E_{l m}$. We observe that the nature of a mode can change if $|\hat\alpha\Delta_{l m}^{\pm}|\sim 1$, meaning that the corrections can push the mode to the other side of the phase boundary. In addition, $\Delta_{l m}^{\pm}$ gives us an estimation of the relative correction to the imaginary part of QNMs close to the phase boundary. In fact, the eikonal computation shows that for DMs close to $\mu_{\rm cr}$, the imaginary part scales with $|E_{l m}|^{3/2}$ --- see \req{omegaIII}. It is reasonable to assume that for finite $l$ modes a similar scaling occurs but replacing $E_{l m}$ by the exact value ${}_{s}E_{l m}$. Therefore, the relative correction to ${}_{s}E_{l m}$ should give us an idea of the relative correction to $\omega_{I}$, roughly $\delta\omega_{I}/\omega_{I}^{\rm Kerr}\sim -3/2 \hat\alpha  \Delta_{l m}^{\pm}$.

\begin{figure}[t!]
  	\centering
  	\includegraphics[width=0.48\textwidth]{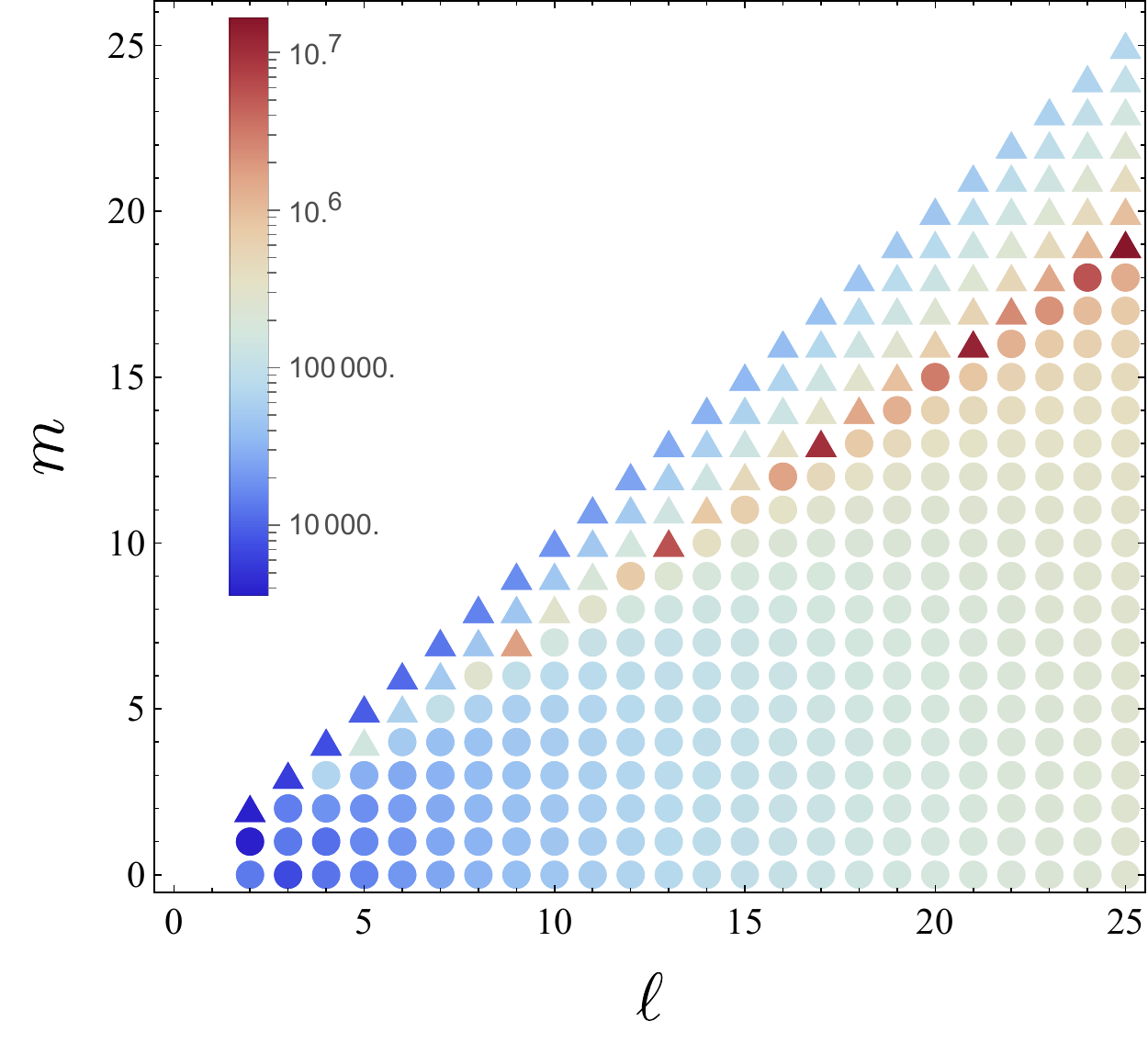}
  	\caption{Coefficient $\Delta_{l m}^{+}$ for the relative corrections to near-extremal QNMs in the theory \req{eq:ISO}. The color of the markers represents the magnitude of $\Delta_{l m}^{+}$ in a logarithmic scale. Triangles are $(l, m)$ values for which only ZDMs exist in GR and circles are values for which DMs  exist.  The most sensitive modes (in darker red) are precisely those on the boundary from one region to another.}
  	\label{fig:lmplane}
 \end{figure}

In Figure~\ref{fig:lmplane} we show the factor $\Delta_{l m}^{+}$ (the case of $\Delta_{l m}^{-}$ is very similar) for all values of $(l, m)$ up to $l=25$. The color of the markers represents the magnitude of $\Delta_{l m}^{+}$ on a logarithmic scale. Circles represent modes with ${}_{s}E_{l m}^{\rm Kerr}<0$ (for which DMs exist in GR) and triangles are modes with ${}_{s}E_{l m}^{\rm Kerr}>0$ (so they are ZDMs in GR). In addition, let us note that $\delta A_{l m}^{\pm}<0$ for all the modes in this particular theory. Therefore, if $\alpha>0$, the corrections tend to reduce the number of DMs (some of the circles would become triangles) and if $\alpha<0$ they tend to increase the number of DMs (triangles near the critical line would become circles).

As we can see in Figure~\ref{fig:lmplane}, there are a few special modes that stand out as they receive much larger corrections than their neighbors.  Some of the most prominent modes are $\Delta_{5,4}^{+}=-1.54\times 10^5$, $\Delta_{9,7}^{+}=-1.74 \times 10^6$, $\Delta_{13, 10}^{+}=-5.71 \times 10^6$, $\Delta_{21, 16}^{+}=-1.29 \times 10^7$.  Therefore, even if $\hat\alpha$ is very small, those modes could be greatly affected by the higher-derivative corrections. All the values of $\Delta_{l m}^{\pm}$ up to $l=13$ can be found in Tables~\ref{table:relcorrectionplus} and \ref{table:relcorrectionminus} in the appendix.

We also remark that, for moderate rotation, the relative corrections to QNM frequencies are of the order of $\hat\alpha$ times an order one constant \cite{Cano:2024ezp}. The values we obtain for $\Delta_{l m}^{\pm}$ show that, in the near-extremal regime, and especially, near to the phase boundary, the corrections become multiple orders of magnitude larger than that.

The computation of $\Delta_{l m}^{\pm}$ can easily be extended to general higher-derivative corrections by using the results of \cite{Cano:2024bhh}, but we expect that the most sensitive modes are universal, since the key factor is how close they lie to the original phase boundary in GR.

 \section{Discussion} \label{sec:discussion}
We have found that the QNMs lying near the phase boundary between DMs and ZDMs are very sensitive to modifications of GR. These modes are long lived in the extremal limit, and we have shown that the relative corrections to their lifetimes are inversely proportional to their distance to the critical point $\bar\mu_{\rm cr}$ --- see \req{relIII}.  The perturbative expansion of higher-derivative corrections to the QNM frequencies breaks down when the corrections become comparable to the GR contribution, but we have shown that this is not necessarily associated to a breakdown of the EFT. Instead, the growth of the correction is associated to the position of the phase boundary being modified. As a consequence, some modes can ``cross'' the boundary, leading to a change in the number of DMs and a qualitative change in the spectrum.  We have shown that this situation can take place in a regime in which the perturbative higher-derivative expansion is under control. That is, a regime in which any additional higher-derivative corrections are subleading and can be neglected. 
This means that, unless another reason prevents us from trusting the theory in that regime, the phenomenon of boundary-crossing is a legitimate prediction of higher-derivative extensions of GR. 
 
One possible reason for not trusting the theory is the presence of new degrees of freedom at a certain scale. We saw that, if the scale of the coupling constant $\alpha\sim \ell_{\rm new}^6$ is the same as the scale of new massive degrees of freedom, the phenomenon of boundary-crossing cannot happen (at least, not for eikonal modes) within the usual regime of validity of the EFT.  This is because it generically requires that the modes have a wavelength shorter than the Compton wavelength $\ell_{\rm UV}$ of the new degrees of freedom. But even if we restrict ourselves to this scenario, the effects of the higher-derivative corrections are still amplified for modes near the phase boundary. In fact, the relative correction to their damping times can be as large as $\mathcal{O}(\hat\alpha^{1/3})$, much larger than the naive order-of-magnitude estimation $\mathcal{O}(\hat\alpha)$. On the other hand, our results indicate that it is enough for the length scale of the higher-derivative corrections to be just slightly above the UV scale, $\ell_{\rm new}\gtrsim\ell_{\rm UV}$, for some modes to cross the phase boundary within the regime of validity of the EFT. 

While most of our analysis has focused on eikonal modes, we have also been able to study the modification of the phase boundary for lower $l$ modes and obtain an estimation for the size of the relative corrections to the modes' damping time. Our results have pointed out a few gravitational modes that are particularly sensitive to corrections, as they lie very close to the boundary: $(5,4)$, $(9,7)$, $(13,10), \ldots$.   The coefficients of the relative correction $\Delta_{l m}^{\pm }$ become very large for those modes, indicating that, even if $\hat\alpha\ll 1$,  large deviation with respect to GR could occur.  All this indicates that the behavior of QNMs near extremality in theories with higher-derivative corrections is likely to be very involved. We still do not have the necessary tools to analyze those QNMs outside the eikonal approximation, but in the light of our results, this question should be addressed in the future. 

Beyond specific considerations, we expect that the sensitivity of the modes near the phase boundary is a general phenomenon that affects any modification of GR or of the Kerr metric. This is because any modification of the Teukolsky equation will generically change the position of the critical point in the potential, hence changing the phase boundary.  For instance, adding a small electric charge could probably also trigger this effect. In fact, the spectrum of near-extremal Kerr-Newman black holes is known to be highly intricate as it exhibits the phenomenon of eigenvalue repulsion \cite{Dias:2021yju,Dias:2022oqm}. It would be interesting to understand if, in the weakly charged limit, this is somehow related to a modification of the phase boundary of DMs and ZDMs \cite{Saha:2025nsg}.
 
Finally, it would be crucial to understand the impact of the corrections to the QNMs on the time-domain waveform. It is known that the QNM spectrum of black holes can suffer from instabilities that nonetheless do not affect the time-domain signal \cite{Daghigh:2020jyk,Jaramillo:2020tuu,Cheung:2021bol}. We do not believe that the effect we have uncovered can be classified as this type of instability and we expect that it will have observational consequences for the ringdown signal. Still, understanding exactly how it is modified is important if we want to test the presence of new physics in ringdown detections.

To sum up, our findings show that, generically, the effects of new physics on the QNM spectrum of black holes become much larger near extremality.  In addition, the ringdown phase of near-extremal black holes is long-lived and, if observed in future experiments, would allow for an exceptionally precise measurement of the corresponding QNM frequencies.  All in all, this implies that black hole spectroscopy of near-extremal black holes would be one of the most promising avenues to test new gravitational physics.  More work is needed on the theoretical front to fully exploit this exciting opportunity. 

\vspace{0.1cm}
\begin{acknowledgments} 
The work of PAC was supported by a fellowship from ``la Caixa'' Foundation (ID 100010434) with code LCF/BQ/PI23/11970032 and by a Ram\'on y Cajal fellowship (RYC2023-044375-I) from Spain’s Ministry of Science, Innovation and Universities.  MD is supported in part by the Research Foundation – Flanders FWO with grants G094523N, G003523N and the Postdoctoral Fellows of the Research Foundation - Flanders grant (1235324N) as well as C16/25/010 from KU Leuven. GvdV is supported by a Proyecto de Consolidación Investigadora (CNS2023-143822) from Spain’s Ministry of Science, Innovation and Universities. This research was carried out in part during a scientific visit funded by the COST action CA22113 ``Fundamental Challenges in
Theoretical Physics."
\end{acknowledgments}

\appendix

\clearpage
\onecolumngrid

\section{Relative correction to the phase boundary condition}\label{App:relcorrectiontables}

We show the values of the relative corrections to the phase boundary condition for both polarities in Tables~\ref{table:relcorrectionplus} and \ref{table:relcorrectionminus}, which  correspond to Figure~\ref{fig:lmplane}. It is clear from the data below that certain modes exhibit very large $\Delta^{\pm}_{lm}$, \textit{e.g.}, $(l,m)=(5,4),(9,7),(13,10)$.

\begin{table*}[h]
	\centering
	\renewcommand\arraystretch{1.2}
	\setlength{\tabcolsep}{2pt}
	\begin{tabularx}{\textwidth}{|p{0.83cm}|*{12}{S|}} 
		\hline
		\diagbox[dir=NW,width=3em,height=2.4em,innerleftsep=4pt,innerrightsep=6pt]{$m$}{$l$} & {2} & {3} & {4} & {5} & {6} & {7} & {8} & {9} & {10} & {11} & {12} & {13} \\ \hline
		\multicolumn{1}{|c|}{0} & 1.34 & 0.70 & 1.23 & 1.50 & 2.01 & 2.55 & 3.19 & 3.91 & 4.71 & 5.59 & 6.55 & 7.59 \\ \hline
		\multicolumn{1}{|c|}{1} & 0.36 & 1.31 & 1.14 & 1.63 & 2.04 & 2.60 & 3.23 & 3.94 & 4.74 & 5.62 & 6.58 & 7.62 \\ \hline
		\multicolumn{1}{|c|}{2} & -0.37 & 1.42 & 1.87 & 1.78 & 2.25 & 2.75 & 3.37 & 4.07 & 4.86 & 5.73
		& 6.69 & 7.73 \\ \hline
		\multicolumn{1}{|c|}{3} & {} & -0.57 & 6.74 & 2.97 & 2.71 & 3.13 & 3.66 & 4.33 & 5.09 & 5.94 & 6.89 & 7.92 \\ \hline
		\multicolumn{1}{|c|}{4} & {} & {} & -0.74 & -15.44 & 5.14 & 4.08 & 4.32 & 4.83 & 5.51 & 6.31 & 7.22 & 8.23 \\ \hline
		\multicolumn{1}{|c|}{5}& {} & {} & {} & -0.91 & -6.37 & 10.08 & 6.18 & 5.95 & 6.32 & 6.97 & 7.78 & 8.72 \\ \hline
		\multicolumn{1}{|c|}{6} & {} & {} & {} & {} & -1.09 & -5.07 & 27.79 & 9.59 & 8.21 & 8.25 & 8.77 & 9.54 \\ \hline
		\multicolumn{1}{|c|}{7}& {} & {} & {} & {} & {} & -1.28 & -4.72 & -173.68 & 15.70 & 11.44 & 10.75 & 10.99 \\ \hline
		\multicolumn{1}{|c|}{8}& {} & {} & {} & {} & {} & {} & -1.48 & -4.68 & -30.12 & 28.88 & 16.24 & 14.08 \\ \hline
		\multicolumn{1}{|c|}{9} & {} & {} & {} & {} & {} & {} & {} & -1.70 & -4.78 & -19.57 & 73.61 & 23.86 \\ \hline
		\multicolumn{1}{|c|}{10} & {} & {} & {} & {} & {} & {} & {} & {} & -1.94 & -4.98 & -16.00 & -570.81 \\ \hline
		\multicolumn{1}{|c|}{11} & {} & {} & {} & {} & {} & {} & {} & {} & {} & -2.20 & -5.23 & -14.38 \\ \hline
		\multicolumn{1}{|c|}{12} & {} & {} & {} & {} & {} & {} & {} & {} & {} & {} & -2.48 & -5.53 \\ \hline
		\multicolumn{1}{|c|}{13}& {} & {} & {} & {} & {} & {} & {} & {} & {} & {} & {} & -2.77 \\ \hline
	\end{tabularx}
	\caption{Values of $\Delta^{+}_{lm}$ for $l=2$ up to $l=13$. Each value should be multiplied by $10^4$.}
	\label{table:relcorrectionplus}
\end{table*}

\begin{table*}[h]
	\centering
	\renewcommand\arraystretch{1.2}
	\setlength{\tabcolsep}{2pt}
	\begin{tabularx}{\textwidth}{|p{0.83cm}*{12}{|S}|}
		\hline
		\diagbox[dir=NW,width=3em,height=2.4em,innerleftsep=4pt,innerrightsep=6pt]{$m$}{$l$} & {2} & {3} & {4} & {5} & {6} & {7} & {8} & {9} & {10} & {11} & {12} & {13} \\ \hline
		\multicolumn{1}{|c|}{0} & 0.15 & 0.85 & 0.98 & 1.49 & 1.94 & 2.52 & 3.16 & 3.88 & 4.69 & 5.57 & 6.53 & 7.58 \\ \hline
		\multicolumn{1}{|c|}{1} & 1.43 & 0.62 & 1.21 & 1.49 & 2.01 & 2.55 & 3.20 & 3.92 & 4.72 & 5.60 & 6.57 & 7.61 \\ \hline
		\multicolumn{1}{|c|}{2} & -0.28 & 3.25 & 1.34 & 1.81 & 2.16 & 2.71 & 3.33 & 4.05 & 4.84 & 5.72 & 6.67 & 7.71 \\ \hline
		\multicolumn{1}{|c|}{3} & {} & -0.46 & 11.41 & 2.54 & 2.71 & 3.06 & 3.63 & 4.30 & 5.07 & 5.93 & 6.87 & 7.90 \\ \hline
		\multicolumn{1}{|c|}{4} & {} & {} & -0.63 & -21.90 & 4.76 & 4.07 & 4.27 & 4.80 & 5.48 & 6.29 & 7.21 & 8.21 \\ \hline
		\multicolumn{1}{|c|}{5} & {} & {} & {} & -0.80 & -8.08 & 9.73 & 6.17 & 5.91 & 6.29 & 6.95 & 7.76 & 8.71 \\ \hline
		\multicolumn{1}{|c|}{6} & {} & {} & {} & {} & -0.98 & -5.99 & 27.39 & 9.58 & 8.18 & 8.22 & 8.75 & 9.52 \\ \hline
		\multicolumn{1}{|c|}{7} & {} & {} & {} & {} & {} & -1.18 & -5.31 & -173.06 & 15.70 & 11.41 & 10.73 & 10.98 \\ \hline
		\multicolumn{1}{|c|}{8} & {} & {} & {} & {} & {} & {} & -1.39 & -5.10 & -30.18 & 28.90 & 16.22 & 14.06 \\ \hline
		\multicolumn{1}{|c|}{9} & {} & {} & {} & {} & {} & {} & {} & -1.62 & -5.10 & -19.67 & 73.72 & 23.85 \\ \hline
		\multicolumn{1}{|c|}{10} & {} & {} & {} & {} & {} & {} & {} & {} & -1.87 & -5.22 & -16.09 & -571.80 \\ \hline
		\multicolumn{1}{|c|}{11} & {} & {} & {} & {} & {} & {} & {} & {} & {} & -2.13 & -5.42 & -14.47 \\ \hline
		\multicolumn{1}{|c|}{12} & {} & {} & {} & {} & {} & {} & {} & {} & {} & {} & -2.41 & -5.69 \\ \hline
		\multicolumn{1}{|c|}{13} & {} & {} & {} & {} & {} & {} & {} & {} & {} & {} & {} & -2.71 \\ \hline
	\end{tabularx}
	\caption{Values of $\Delta^{-}_{lm}$ for $l=2$ up to $l=13$. Each value should be multiplied by $10^4$.}
	\label{table:relcorrectionminus}
\end{table*}

\clearpage

\twocolumngrid

\bibliography{Gravities}
\noindent \centering

\end{document}
%